\documentclass[aps,superscriptaddress,amsmath,amssymb,twocolumn,showpacs,floatfix,english,noshowpacs]{revtex4-2}

\usepackage{url}
\usepackage{bm,bbm}
\usepackage{graphicx}
\usepackage{color}
\usepackage[colorlinks=true, urlcolor=blue, linkcolor=blue, citecolor=blue, pdftex]{hyperref}
\usepackage[english]{babel}

\usepackage{soul}
\usepackage{blindtext}
\usepackage{lipsum}
\usepackage{nicefrac}

\newcommand{\dagga}{{\phantom{\dagger}}}
\newcommand{\dis}{\displaystyle}
\newcommand{\fract}[2]{\frac{\dis #1}{\dis #2}}
\newcommand{\esp}[1]{\text{e}^{#1}}
\newenvironment{eqs}%
{\begin{equation} \begin{aligned}}%
{\end{aligned} \end{equation} }
\newcommand{\beal}{\begin{eqs}}
\newcommand{\eal}{\end{eqs}}

\begin{document}

\title{Insulating and metallic phases in the one-dimensional Hubbard-Su-Schrieffer-Heeger model: Insights from a backflow-inspired variational wave function}

\author{Davide Piccioni}
\email[]{dpiccion@sissa.it}
\affiliation{Scuola Internazionale Superiore di Studi Avanzati (SISSA), Via Bonomea 265, I-34136, Trieste, Italy}

\author{Francesco Ferrari}
\affiliation{Institute for Theoretical Physics, Goethe University Frankfurt, Max-von-Laue-Stra{\ss}e 1, D-60438 Frankfurt am Main, Germany}

\author{Michele Fabrizio}
\affiliation{Scuola Internazionale Superiore di Studi Avanzati (SISSA), Via Bonomea 265, I-34136, Trieste, Italy}

\author{Federico Becca}
\affiliation{Dipartimento di Fisica, Universit\`a di Trieste, Strada Costiera 11, I-34151 Trieste, Italy}

\date{\today}

\begin{abstract}
The interplay between electron-electron and electron-phonon interactions is studied in a one-dimensional lattice model, by means of a variational Monte Carlo
method based on generalized Jastrow-Slater wave functions. Here, the fermionic part is constructed by a pair-product state, which explicitly depends on the
phonon configuration, thus including the electron-phonon coupling in a backflow-inspired way. We report the results for the Hubbard model in presence of the
Su-Schrieffer-Heeger coupling to optical phonons, both at half-filling and upon hole doping. At half-filling, the ground state is either a translationally
invariant Mott insulator, with gapless spin excitations, or a Peierls insulator, which breaks translations and has fully gapped excitations. Away from
half-filling, the charge gap closes in both Mott and Peierls insulators, turning the former into a conventional Luttinger liquid (gapless in all excitation
channels). In the latter case, instead, a finite spin gap remains at small doping. Even though consistent with the general
theory of interacting electrons in one dimension, the existence of such a phase (with gapless charge but gapped spin excitations) has never been demonstrated
in a model with repulsive interaction and with only two Fermi points. Since the spin-gapped metal represents the one-dimensional counterpart of a superconductor,
our results furnish evidence that a true off-diagonal long-range order may exist in the two-dimensional case.
\end{abstract}

\maketitle

\section{Introduction}\label{sec:intro}

The electron-phonon interaction plays an important role in condensed-matter physics, as it often leads to stark modifications of the low-energy electronic properties,
the most striking example being the insurgence of superconductivity in a variety of materials~\cite{degennes}. In addition, it may favor lattice deformations, turning,
e.g., a metallic system into a (Peierls) insulator~\cite{peierls}. In molecular solids, the electron-phonon coupling is at the origin of the Jahn-Teller effect, which
can lead to orbital order and geometric distortions~\cite{khomskii}. In high-temperature (cuprate) superconductors, the interplay between electron pairing and
charge-density waves (the so-called stripes) has been widely discussed in the past 30 years and represents a crucial aspect that must be elucidated in order to reach
a definitive understanding of those materials~\cite{kivelson2003,fradkin2015}.

From a computational perspective, an {\it ab-initio} technique capable of accurately treating both electron-electron (e-e) and electron-phonon (e-ph) interactions
is currently lacking, especially when effects beyond the Born-Oppenheimer approximation are relevant. In this respect, the definition of suitable low-energy models
on a lattice may serve as a valuable framework for shedding light on the phenomena that emerge in this kind of problems. The single-band Hubbard model (or its
strong-coupling version, the $t-J$ model) serves as the simplest example to capture electron correlations~\cite{arovas2022,qin2022}. Within this approach, phonons
can be included by adding harmonic oscillators on each lattice site, leading to acoustic and/or optical branches. Then, the e-ph coupling can be modeled by Fr\"ohlich,
Holstein, or Su-Schrieffer-Heeger (SSH) Hamiltonians~\cite{frohlich1954,holstein1959,su1979,su1980,hohenadler2018}. The former two cases are suitable to describe
polaron effects in dielectric crystals; the latter one has been introduced to characterize solitons in one-dimensional materials (such as Polyacetylene). Here,
lattice vibrations are directly coupled to the electron hopping.

In one spatial dimension, whenever the electron band is half filled and in the absence of e-e interaction, the SSH model gives rise to a Peierls instability -- a
spontaneous dimerization of the chain, with alternating long and short bonds between nearest-neighbor sites -- as obtained by earlier quantum Monte Carlo calculations
and renormalization-group arguments~\cite{hirsch1983a}. In the presence of an e-e interaction, such as the onsite Hubbard repulsion, the situation is more complicated.
In the adiabatic limit (i.e., when taking an infinite ion mass and, therefore, no ion dynamics), lattice distortions appear for infinitesimal e-ph
coupling~\cite{baeriswyl1985}, as also previously found in the strong-coupling (Heisenberg) model~\cite{cross1979}. A full quantum mechanical treatment of the model,
including both electron and ion dynamics, has been addressed in a relatively limited number of works. In Ref.~\cite{hirsch1983b}, it has been argued that an on-site
electron repulsion suppresses the Peierls distortions for large phonon frequencies. Later, a few works, mainly using Monte Carlo or density-matrix renormalization
group (DMRG) techniques, analyzed a variety of models with both quantum lattice fluctuations and short-range e-e interactions~\cite{sengupta2003,pearson2011,assaad2015}.
In particular, acoustic and optical phonons have been compared at half-filling (also considering different ways to couple lattice vibrations to the electronic
hopping)~\cite{costa2023}. The general outcome suggests that two phases can be stabilized: the Mott insulator (when the e-e interaction dominates), with no lattice
distortions and gapless spin excitations, and the Peierls insulator (when the e-ph coupling dominates), characterized by lattice distortions, a two-fold degenerate
ground state, and a fully gapped spectrum. The transition between them is described by the Kosterliz-Thouless universality class~\cite{giamarchi}, making its precise
location extremely challenging.

Away from half-filling, a limited number of investigations have been attempted. Indeed, ground-state Monte Carlo techniques suffer from the sign problem, which
prevents one from assessing large system sizes, and the DMRG accuracy is strongly affected by the large entanglement of the gapless (metallic) ground state. One
interesting aspect is to understand whether a so-called Luther-Emery phase~\cite{luther1974}, namely a metal with gapped spin excitations, may ever emerge when
doping the Peierls insulator. In this respect, analytic and numerical works demonstrated that the Luther-Emery liquid appears in the doped two-leg ladder Hubbard
model (with no phonons)~\cite{fabrizio1996,daul1998,nishimoto2008,balents1996,shen2023}. Its existence is triggered by the presence of multiple Fermi points in the
non-interacting band structure, as discussed in weak-coupling approaches~\cite{fabrizio1996,balents1996}. For example, in the one-dimensional Hubbard model with
both nearest- ($t$) and next-nearest-neighbor ($t^\prime$) hopping, a metallic phase with gapped spin excitations emerges by lightly doping the dimerized insulator
found at $t^\prime>t/2$~\cite{fabrizio1996,daul1998,nishimoto2008}. 
In addition, Ref.~\cite{sandvik2002} suggested that the Luther-Emery liquid may also emerge from
doping an insulating phase with bond order obtained within a single-band Hubbard model in presence of both on-site and nearest-neighbor interactions.
By analogy, one could expect that the same phase could appear upon doping the (dimerized) Peierls
insulator stabilized by the electron-phonon interaction. An early work~\cite{yonemitsu1996}, based upon bosonization and renormalization-group methods, indeed showed
that a spin-gapped metal emerges close to half-filling due to the phonon-assisted backward scattering. More recently, both static and dynamical properties have
been addressed by DMRG~\cite{banerjee2023}. However, a single electron density $n$ has been considered, relatively far away from half-filling (i.e., $n=0.75$), and
Luttinger-liquid properties are observed, with gapless charge and spin degrees of freedom.

In this work, we study the one-dimensional Hubbard-SSH model by means of a variational Monte Carlo approach (VMC). We employ a variational wave function composed
of three terms: the first one is a simple phonon condensate, which can give rise to a staggered (dimerized) pattern of the lattice displacements; the second one
is a Bardeen-Cooper-Schrieffer (BCS) state; the last one is a Jastrow factor, which includes electron-electron, phonon-phonon, and electron-phonon correlations.
The main novelty of the variational {\it Ansatz} resides in the inclusion of ``backflow'' correlations within the BCS wave function, through an auxiliary BCS
Hamiltonian which explicitly depends upon the phonon coordinates. We present a detailed analysis of the SSH model in presence of a local Hubbard interaction, both
at half-filling and for various hole doping. At half-filling, the ground state is either a Mott or a Peierls insulator, depending on the values of the e-e and e-ph
couplings, in agreement with previous calculations~\cite{sengupta2003,pearson2011,assaad2015}. The central part of our work concerns the doped case, where metallic
states appear. In fact, while doping a Mott insulator gives rise to a standard Luttinger liquid, similarly to what happens in the Hubbard model without phonons,
in a lightly doped Peierls insulator the spin gap remains finite, thus stabilising a Luther-Emery liquid, which has never been noticed in this kind of coupled
electron-phonon problems.

The rest of the paper is structured as follows. In Sec.~\ref{sec:method}, we introduce the model and discuss the form of the variational {\it Ansatz}, with backflow
correlations. In Sec.~\ref{sec:results}, we give a benchmark for the accuracy of our wave function against DMRG results, then we show the results at half-filling
and upon doping; finally, in Sec.~\ref{sec:concl}, we draw our conclusions and perspectives for future works.

\section{Model and method}\label{sec:method}

\subsection{The electron-phonon Hamiltonian}

We consider a one-dimensional electron-phonon system in which, besides the onsite Hubbard interaction, the electron hopping is modulated by the phonon displacements
and harmonic oscillators are located on each site of the lattice (modeling optical phonons):
\begin{eqnarray}
\hat{\cal H} &=& -t\sum_{i,\sigma} \left[1-\alpha\left(\hat{x}_{i+1}-\hat{x}_{i}\right)\right] \hat{c}^\dag_{i,\sigma} \hat{c}^\dagga_{i+1,\sigma} + {\rm h.c.} \nonumber \\
&+& U\sum_i \hat{n}_{i,\uparrow}\hat{n}_{i,\downarrow}  + \sum_i\left( \frac{\hat{p}_i^2}{2m} +\frac{1}{2}m\omega^2 \hat{x}_i^2 \right).
\label{eq:Hamiltonian1}
\end{eqnarray}
The operator $\hat{c}^\dag_{i,\sigma}$ ($\hat{c}^\dagga_{i,\sigma}$) creates (destroys) an electron on site $i$ with spin $\sigma$,
$\hat{n}_{i,\sigma}=\hat{c}^\dag_{i,\sigma} \hat{c}^\dagga_{i,\sigma}$ is the density (per spin) at site $i$, and $\hat{x}_{i}$ and $\hat{p}_{i}$ are the position
and momentum operators of the phonons, which satisfy the usual commutation relation:
\begin{equation}
\left[\hat{x}_j, \hat{p}_k\right]=i\hbar \delta_{jk}.
\end{equation}
Here, we define dimensionless displacement and momentum operators by the following canonical transformation:
\begin{equation}
\hat{X}_i= \sqrt{\frac{m\omega}{\hbar}} \hat{x}_i \qquad \qquad \hat{P}_i= \sqrt{\frac{1}{m\omega\hbar}} \hat{p}_i,
\end{equation}
such that
\begin{equation}
\left[\hat{X}_i, \hat{P}_i\right]=i.
\end{equation}
Therefore, the Hamiltonian becomes:
\begin{eqnarray}
\hat{\cal H} &=& -t\sum_{i,\sigma} \left[1-\tilde{\alpha} \left(\hat{X}_{i+1}-\hat{X}_{i}\right)\right] \hat{c}^\dag_{i,\sigma} \hat{c}^\dagga_{i+1,\sigma} + {\rm h.c.} \nonumber  \\
&+& U\sum_i \hat{n}_{i,\uparrow}\hat{n}_{i,\downarrow} + \frac{\hbar \omega}{2} \sum_i \left[ \hat{P}_i^2 + \hat{X}_i^2 \right],
\label{eq:Hamiltonian2}
\end{eqnarray}
where $\tilde{\alpha}=\alpha\sqrt{ \frac{\hbar}{m\omega}}$ is the rescaled, dimensionless e-ph coupling. An important (dimensionless) parameter to quantify the e-ph
interaction is
\begin{equation}\label{eq:lambda}
\lambda = \fract{t \alpha^2}{m\omega^2} = \fract{t\tilde{\alpha}^2}{\hbar \omega}
\end{equation}
After this procedure, the mass of the phonons has been completely reabsorbed. In the following, we will take $t$ the energy scale and fix $\hbar \omega=t$. Different
values of the phonon energy have also been considered, without affecting the qualitative results.

\begin{figure}
\includegraphics[width=\columnwidth]{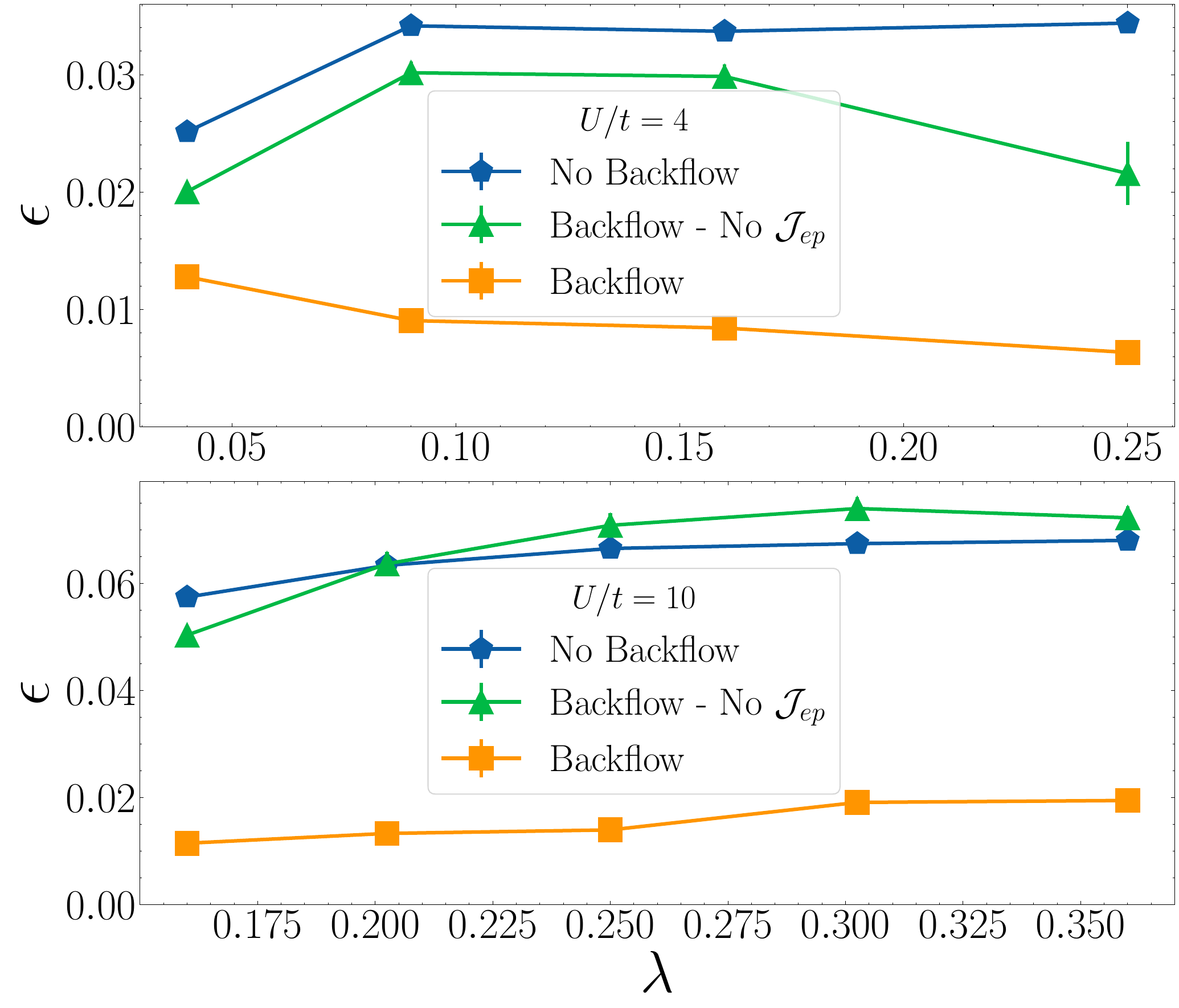}
\caption{\label{fig:energy}
Accuracy of the variational energy with respect to DMRG results $\epsilon=(E-E_{\rm DMRG})/|E_{\rm DMRG}|$ on $L=50$ sites. The simplest wave function, for which
the electronic part is obtained from the auxiliary Hamiltonian~\eqref{eq:BCS}, is denoted by blue points; the best one with backflow correlations generated by
Eq.~\eqref{eq:BCSback}, is denoted by orange points; an intermediate case, with backflow correlations, but without electron-phonon Jastrow factor ${\cal J}_{\rm ep}$,
is also reported for comparison (green points). Results are shown for $U/t=4$ (upper panel) and $U/t=10$ (lower panel), as a function of $\lambda$.}
\end{figure}

\subsection{The variational wave function}

In order to get insights into the ground-state properties of the Hubbard-SSH model of Eq.~\eqref{eq:Hamiltonian2}, we define a variational {\it Ansatz} that is the
product of Jastrow factors (${\cal J}_{\rm ee}$, ${\cal J}_{\rm pp}$, and ${\cal J}_{ep}$), a phonon state ($|\Psi_{\rm p}\rangle$), and an electronic state that
depends parametrically on the phonon displacements ($|\Psi_{\rm e}\rangle$):
\begin{equation}\label{eq:psi0}
|\Psi_{\rm var}\rangle = {\cal J}_{\rm ee} {\cal J}_{\rm pp} {\cal J}_{\rm ep} |\Psi_{\rm p}\rangle \otimes |\Psi_{\rm e}\rangle.
\end{equation}
The variational energy $E_{\rm var}$ is evaluated by performing a Markov chain in the Hilbert space with both electron and phonon configurations
$|X;n_{\sigma}\rangle = |X\rangle \otimes |n_{\sigma}\rangle = \bigotimes_j |X_{j}\rangle \otimes |n_{j,\sigma}\rangle$, i.e., in the local eigenbasis of the
$\hat{X}_{j}$ (phonon) and $\hat{n}_{j,\sigma}$ operators on each lattice site $j$. Specifically,
\begin{align}\label{eq:varener}
E_{\rm var} &= \frac{\langle \Psi_{\rm var}| \hat{\cal H} |\Psi_{\rm var}\rangle}{\langle \Psi_{\rm var}|\Psi_{\rm var}\rangle} =\nonumber \\
&= \sum_{n_{\sigma}} \int d X \, \underbrace{\frac{|\langle X;n_{\sigma}|\Psi_{\rm var}\rangle|^2}{\langle \Psi_{\rm var}|\Psi_{\rm var}\rangle}}_{{\cal P}(X;n_{\sigma})} \,
\underbrace{\frac{\langle X;n_{\sigma}|\hat{\cal H}|\Psi_{\rm var}\rangle}{\langle X;n_{\sigma}|\Psi_{\rm var}\rangle}}_{e_L(X;n_{\sigma})}\;,
\end{align}
where the sum extends over all electron configurations and the integral is over all phonon displacements. Along the Markov process, a set of configurations
$(X; n_{\sigma})_m$ (with $m=1,\dots,M$) are drawn according to the probability ${\cal P}(X;n_{\sigma})$ by using the Metropolis algorithm, which allows us to
estimate the variational energy as:
\begin{equation}
E_{\rm var} \approx \frac{1}{M} \sum_{m=1}^M e_L(X;n_{\sigma})_m.
\end{equation}
In particular, the sampling procedure includes both electron and phonon moves; within the former, either single or double (spin flip) electron hopping at nearest-neighbor sites are considered; the latter consists in local updates of the displacements $X_j \mapsto X_j +\Delta$, with $\Delta$ uniformly distributed  within the interval $[-\Delta_{\rm max},\Delta_{\rm max}]$. The acceptance probability is then computed by the Metropolis algorithm.

Let us now discuss the specific construction of the variational wave function that is used in this work. In previous works~\cite{watanabe2015,karakuzu2017,ferrari2024},
the electronic part has been taken as a pair-product state:
\begin{equation}
|\Psi_{\rm e}\rangle = \exp \left( \sum_{i,j} f_{i,j} \hat{c}_{i,\uparrow}^{\dagger} \hat{c}_{j,\downarrow}^{\dagger} \right ) |0\rangle,
\end{equation}
which can be obtained from the ground state of an auxiliary BCS Hamiltonian:
\begin{eqnarray}
\hat{\cal H}_0 &=& \sum_{i,j,\sigma} t_{ij} \hat{c}^\dag_{i,\sigma} \hat{c}^\dagga_{j,\sigma} + \mu \sum_{i,\sigma} \hat{c}^\dag_{i,\sigma} \hat{c}^\dagga_{i,\sigma} \nonumber \\
&+& \sum_{i,j} \Delta_{ij} \hat{c}^\dag_{i,\uparrow} \hat{c}^\dag_{j,\downarrow} + \textit{h.c.},
\label{eq:BCS}
\end{eqnarray}
where the hopping ($t_{ij}=t^{*}_{ji}$) and pairing ($\Delta_{ij}=\Delta_{ji}$) amplitudes are variational parameters, as well as the chemical potential $\mu$.
Alternatively, it is possible to optimize directly the pairing function $f_{i,j}$, without passing through the definition of the auxiliary Hamiltonian~\cite{ohgoe2017}.
Within these approaches, the electronic wave function does not depend upon the phonon displacements, e.g., the pairing function does not change along the Markov chain,
once the variational parameters are kept fixed.

Here, we generalize this construction, by taking an auxiliary Hamiltonian that parametrically depends upon the phonon displacements:
\begin{eqnarray}
\hat{\cal H}_{\rm ep} &=& \hat{\cal H}_0 + \sum_{i,m,\sigma} g_{m} \left(X_{i+m}-X_{i}\right) \hat{c}^\dag_{i,\sigma} \hat{c}^\dagga_{i+m,\sigma} \nonumber \\
&+& \sum_{i,m} h_{m} \left(X_{i+m}-X_{i}\right) \hat{c}^\dag_{i,\uparrow} \hat{c}^\dag_{i+m,\downarrow} + \textit{h.c.},
\label{eq:BCSback}
\end{eqnarray}
where $g_{m}=g_{-m}$ and $h_{m}=-h_{-m}$ are variational parameters for $m=\pm 1$ and $\pm 3$ (suitable for a possible bond dimerization) and $\{ X_j \}$ are the
phonon displacements in the configuration $|X\rangle$ visited along the Markov chain. In this way, $\langle X;n_{\sigma}|\Psi_{\rm e}\rangle$ depends upon the actual
phonon configuration. In this sense, a sort of ``backflow'' correlations are included in the electronic part of the variational wave function. In the standard case
on the continuum, the effective position of every electron (from which the Slater determinant is constructed) depends on all the other
ones~\cite{feynman1956,lee1981,schmidt1981}; on the lattice, this approach has been extended by constructing single-particle orbitals or pairing functions that
explicitly depend upon the many-body electron configuration~\cite{tocchio2008,tocchio2011}. Here, instead, backflow correlations involve electrons and phonons.

The phonon part has the same form as used in Ref.~\cite{ferrari2024}:
\begin{equation}\label{eq:phonon}
\langle X|\Psi_{\rm p}\rangle = \prod_{j}\exp{\left\{ -\frac{1}{2}\left[ X_{j} - (-1)^j z \right]^2 \right\}},
\end{equation}
where the single variational parameter ($z$) determines the staggered phonon displacement.

Within the simple approach, where the electronic state is obtained from the BCS Hamiltonian~\eqref{eq:BCS} (i.e., without backflow correlations), the Peierls
instability is also accompanied by a breaking of the translational symmetry in the hopping $t_{ij}$ and pairing $\Delta_{ij}$ parameters, which may give a
considerable energy gain (with respect to the uniform case), when the phonon parameter $z$ becomes finite. The staggered intensity of hopping (and pairing) is
necessary to get the correct periodicity in the electronic correlations. By contrast, within the extended framework of the auxiliary Hamiltonian~\eqref{eq:BCSback},
it is not necessary to break the translational symmetry in the electronic part of the wave function to obtain accurate correlation functions.

Finally, the e-e and e-ph correlations are included by standard Jastrow factors~\cite{ohgoe2017,ferrari2024}:
\begin{eqnarray}
{\cal J}_{\rm ee} &=& \exp{\left\{ \frac{1}{2} \sum_{i,j} v_{ij} (n_{i}-1)(n_{j}-1) \right\}}, \\
{\cal J}_{\rm pp} &=& \exp{\left\{ \frac{1}{2} \sum_{i,j} u_{ij} X_{i} X_{j} \right\}}, \\
{\cal J}_{\rm ep} &=& \exp{\left\{ \sum_{i,j} w_{ij} (n_{i}-1)(n_{j}-1)(X_{i}-X_{j}) \right\}},\hspace{0.4cm}
\label{eq:jastrowep}
\end{eqnarray}
where the pseudo-potentials $v_{ij}$, $u_{ij}$, and $w_{ij}$ are treated as variational parameters. It turns out that the optimal energy is obtained by imposing both
translation and reflection symmetries, with $v_{ji}=v_{ij}$, $u_{ji}=u_{ij}$, and $w_{ji}=-w_{ij}$. The density-density Jastrow factor is particularly relevant to
describe the Mott insulating phase, in which the Fourier transform of the pseudo-potential $v_{q} \approx 1/q^2$ for small momenta $q$~\cite{capello2005}.

In the following, we consider clusters with $L$ sites and periodic-boundary conditions. All the variational parameters of the variational wave function are optimized
by using the stochastic reconfiguration approach~\cite{sorella2005,becca2017}. Our variational wave function is implemented in a JAX-based code~\cite{jax} that runs
on parallel CPUs thanks to mpi4jax~\cite{mpi4jax}.

\begin{figure}
\includegraphics[width=\columnwidth]{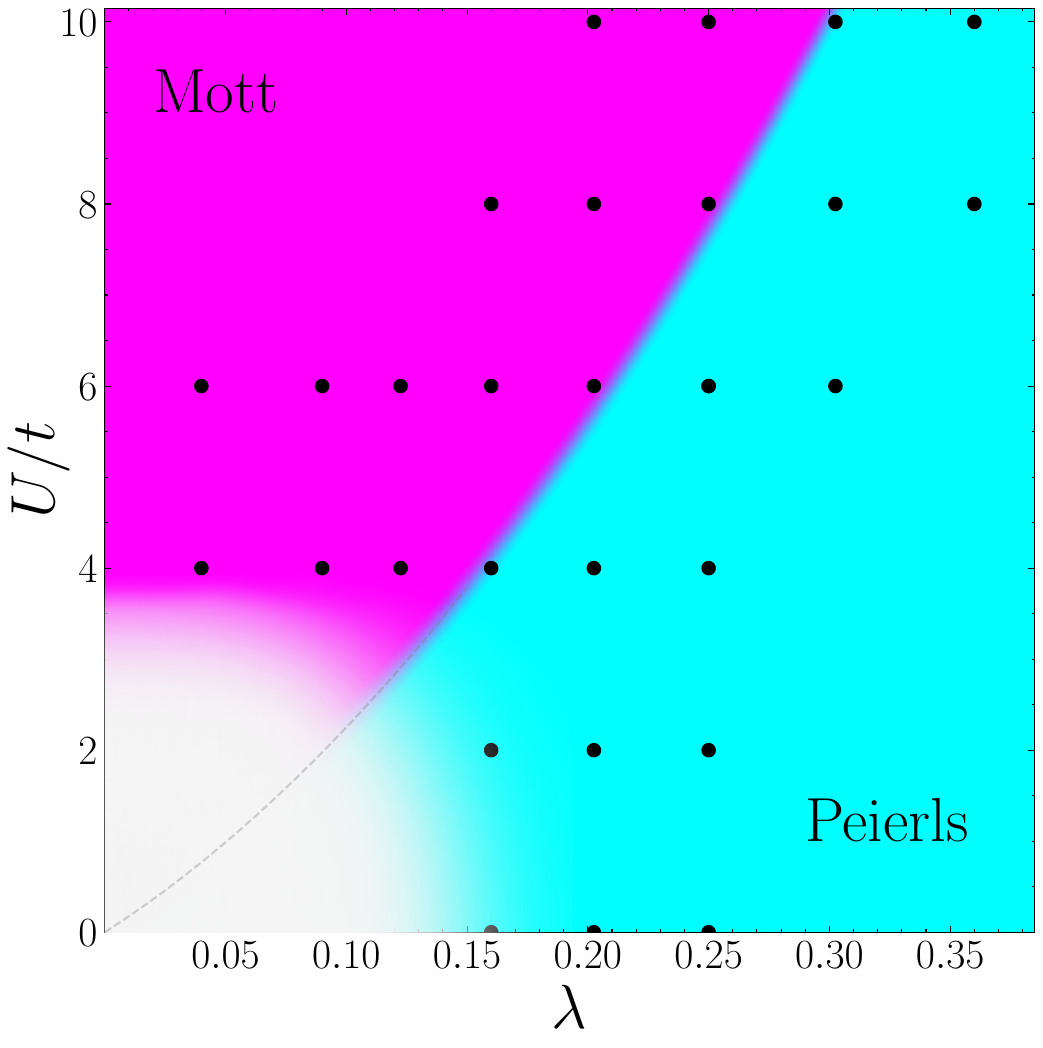}
\caption{\label{fig:phasdiag}
Phase diagram at half-filling as a function of $U/t$ and $\lambda$ with $\hbar \omega/t=1$. Calculations are performed on a chain with $L=50$ sites on the points
marked by filled circles. For $U/t \to 0$ and $\lambda \to 0$,  distinguishing between Mott and Peierls insulators becomes extremely difficult; this fact is
denoted by the white region.}
\end{figure}

\begin{figure*}
\includegraphics[width=0.9\textwidth]{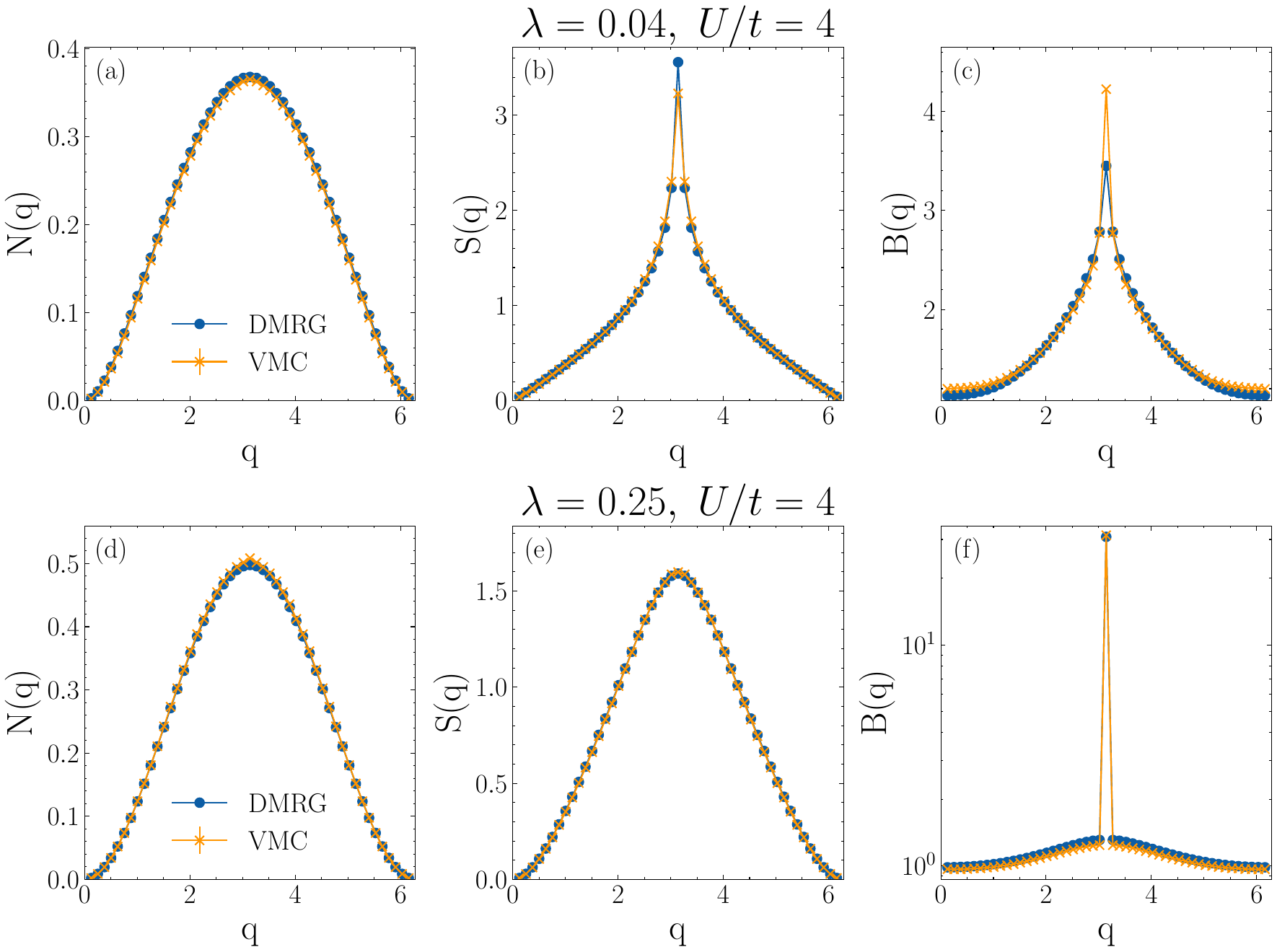}
\caption{\label{fig:compare_Mott_Peierls}
Density, spin, and bond structure factors [denoted by $N(q)$, $S(q)$, and $B(q)$, respectively] at half-filling for $L=50$ sites for $U/t=4$ and $\lambda=0.04$ (upper
panels) and $\lambda=0.25$ (lower panels). The phonon energy is taken $\hbar \omega/t=1$. In each panel, the variational results of the backflow wave function are
compared to the one obtained by DMRG calculations (optimized for the Hamiltonian with periodic-boundary conditions).}
\end{figure*}

\section{Results}\label{sec:results}

\subsection{Benchmarks with DMRG calculations}

First, we would like to assess the accuracy of our approach and compare the variational energies to DMRG results at half-filling, for $L=50$. In the DMRG calculations,
periodic-boundary conditions are taken on the Hamiltonian, but a truncation in the phonon Hilbert space is necessary. Here, we fix a maximum occupancy of $10$ bosons
per site (which is always much larger than the average occupation). Then, the energy of the electron-phonon Hamiltonian is optimized by means of the DMRG algorithm implemented in the ITensor library~\cite{ITensor} for a matrix-product state (MPS) which,
for computational efficiency, has no translational symmetry. The accuracy of DMRG
calculations is verified by evaluating the variance of the total energy, which is always below $0.007t^2$; the bond
dimension of the MPS is always above $\chi=1200$, in some cases reaching $\chi=2400$.

The results of the energy accuracy $\epsilon=(E_{\rm VMC}-E_{\rm DMRG})/|E_{\rm DMRG}|$ for $U/t=4$ and $10$ are reported in Fig.~\ref{fig:energy}, by varying the e-ph
coupling $\lambda$. The best wave function, with backflow correlations and Jastrow factors, reaches a rather high accuracy ($\epsilon \approx 0.01$ and $\approx 0.02$,
for $U/t=4$ and $10$, respectively), largely improving on the results obtained by the standard wave function without backflow terms. Remarkably, the e-ph Jastrow
factor~\eqref{eq:jastrowep} plays an important role in the variational optimization, since a substantial worsening in the energy is detected when removing it (the
accuracy of the backflow state without ${\cal J}_{\rm ep}$ is similar to the one of the standard {\it Ansatz} with no backflow, but including ${\cal J}_{\rm ep}$).

\begin{figure*}
\includegraphics[width=0.9\textwidth]{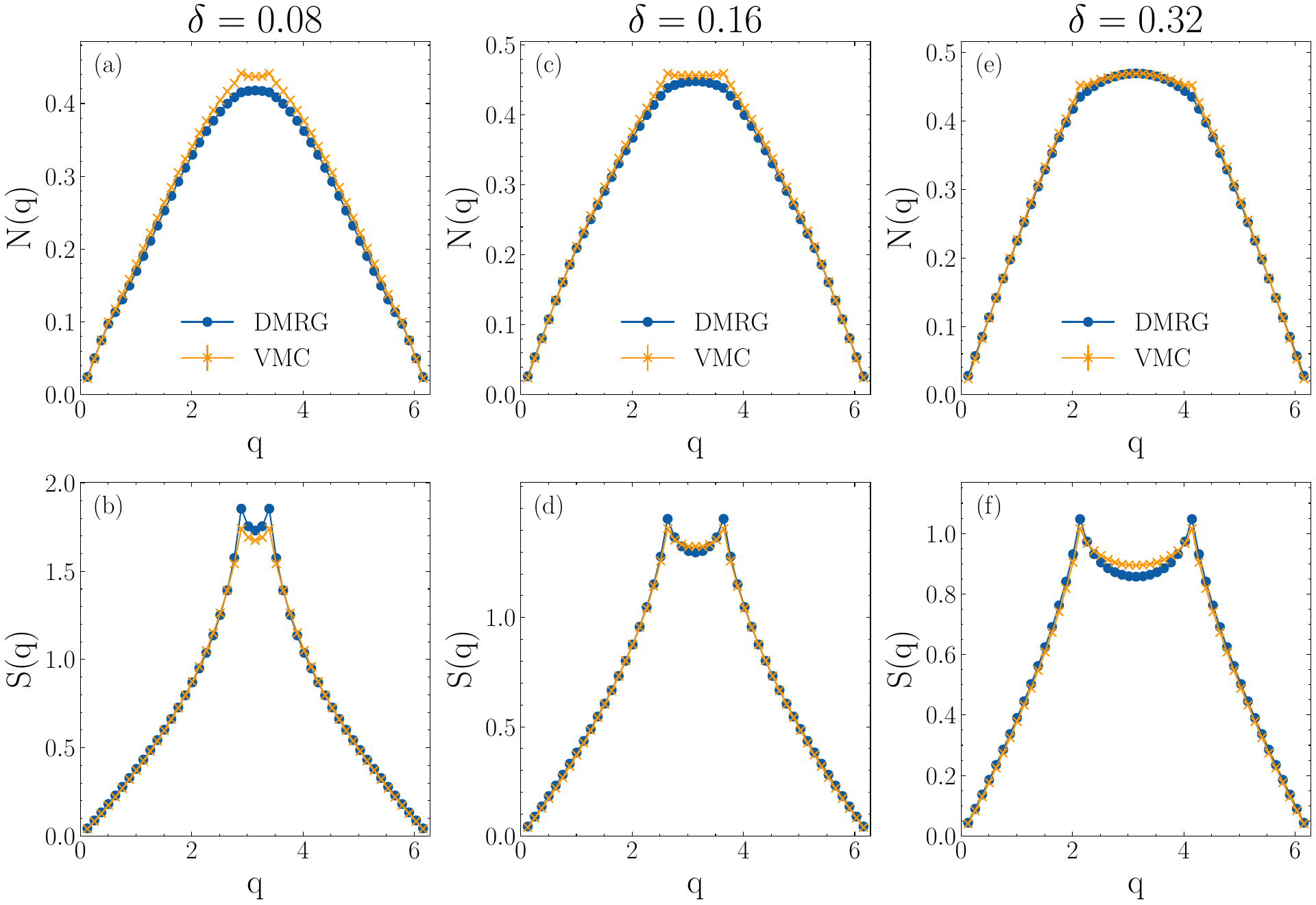}
\caption{\label{fig:doped_Mott}
Density and spin structure factors for a chain with $L=50$ sites at various dopings $\delta$, for $U/t=4$ and $\lambda=0.04$ (with $\hbar \omega/t=1$). At half-filling,
i.e. $\delta=0$, the ground state is a Mott insulator. The results obtained by using DMRG (optimized for the Hamiltonian with periodic-boundary conditions) are also
shown for comparison.}
\end{figure*}

\begin{figure*}
\includegraphics[width=0.9\textwidth]{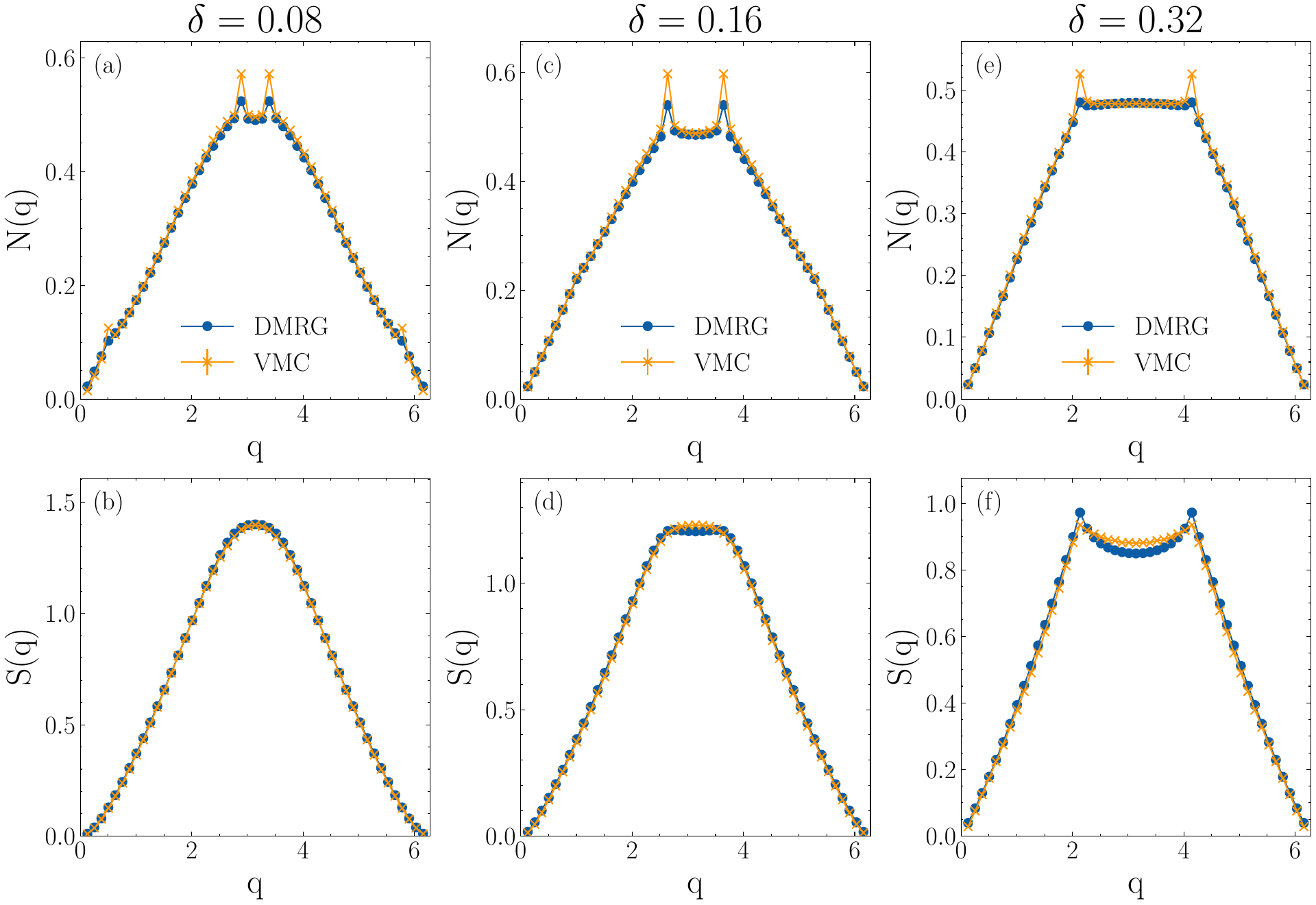}
\caption{\label{fig:doped_Peierls}
The same as in Fig.~\ref{fig:doped_Mott} but for $U/t=4$, $\lambda=0.25$. At half-filling the ground state is a Peierls insulator.}
\end{figure*}

\subsection{Results at half-filling}

Let us now discuss the results at half-filling by focusing on the best variational {\it Ansatz} with backflow correlations. By varying the e-e repulsion $U$ and the
e-ph interaction $\lambda$, the ground state is either a Mott (uniform) or a Peierls (dimerized) insulator~\cite{sengupta2003}. Within our approach, the presence of
a finite lattice distortion is signaled by the stabilization of a finite $z$ parameter in Eq.~\eqref{eq:phonon}, which gives a dimerized pattern around which the
phonon displacements are distributed. In this case, the translational symmetry is explicitly broken in the variational wave function and dimerization can be also
detected from the calculation of the bond-order parameter:
\begin{equation}\label{eq:order_parameter}
B_{\rm e} = \frac{1}{L} \sum_{j=1}^L\, \sum_{\sigma}\,(-1)^j \,\langle \,\hat{c}^\dag_{j,\sigma} \hat{c}^\dagga_{j+1,\sigma} + \hat{c}^\dag_{j+1,\sigma} \hat{c}^\dagga_{j,\sigma}\, \rangle,
\end{equation}
where the expectation value is taken over the variational state $|\Psi_{\rm var} \rangle$.

The results are summarized in the phase diagram shown in Fig.~\ref{fig:phasdiag}, fixing $\hbar \omega/t=1$. Here, the trivial limits are $\lambda=0$, which corresponds
to the Mott insulator with no e-ph coupling, and $U=0$, which corresponds to the Peierls insulator. The transition between these two phases is of the Kosterliz-Thouless
type~\cite{giamarchi}. Therefore, it is extremely difficult to locate its actual location within numerical calculations, since the spin gap of the Peierls phase is
exponentially small close to the transition. This is particularly relevant in the vicinity of the non-interacting limit $U=\lambda=0$, where large clusters become
necessary to detect the presence of the tiny dimerization that exists for $\lambda \to 0$. Nevertheless, sufficiently away from the transition line (and the
non-interacting limit), our variational approach can straightforwardly distinguish two different regimes, with and without bond order.

In addition, the nature of the ground state may be inferred from the equal-time correlation functions:
\begin{equation}\label{eq:structure_factors}
\mathcal{G}(q) = \frac{1}{L} \sum_{m,n} \,\esp{iq(m-n)} \;\langle \hat{O}_{m} \hat{O}_{n} \rangle\,;
\end{equation}
here, $\hat{O}_{m}$ stands for density, spin, or bond operators at site $m$:
\begin{eqnarray}
\hat{n}_{m} &=& \sum_{\sigma} \hat{c}^\dag_{m,\sigma}\hat{c}^\dagga_{m,\sigma}, \\
\hat{S}^z_{m} &=& \sum_{\sigma} s_{\sigma} \hat{c}^\dag_{m,\sigma}\hat{c}^\dagga_{m,\sigma}, \\
\hat{b}_{m} &=& \sum_{\sigma} \left(\hat{c}^\dag_{m,\sigma}\hat{c}^\dagga_{m+1,\sigma} + \text{h.c.} \right),
\end{eqnarray}
where $s_{\sigma}=1$ for spin up and $-1$ for spin down. In the following, the three structure factors will be denoted by $\mathcal{G}=N$, $S$, and $B$ (for density,
spin, and bond operators, respectively). Power-law correlations in real space imply cusps in momentum space and the existence of gapless excitations in the
corresponding sector; instead, an exponential decay gives a smooth behavior in momentum space and gapped excitations.

The results for $U/t=4$ are reported in Fig.~\ref{fig:compare_Mott_Peierls} for two different regimes, i.e., $\lambda=0.04$ (Mott insulator) and $\lambda=0.25$ (Peierls
insulator). The DMRG results (on the same cluster size, optimized for the Hamiltonian with periodic-boundary conditions) are also reported for comparison. First of all,
the difference between Mott and Peierls states appears in the bond structure factor, which shows a huge peak at $q=\pi$ (i.e., $B(\pi) \approx 30$) in the latter case;
the presence of a true dimerization implies a divergence of $B(\pi)$ with $L$. Within the variational approach, the huge peak at $q=\pi$ comes from the stabilization of
a finite phonon parameter $z$ in Eq.~\eqref{eq:phonon}, i.e. a finite lattice distortion. In other words, the existence of a Peierls phase can be more easily detected
by looking at $z$ [or equivalently at $B_{\rm e}$ defined in Eq.~\eqref{eq:order_parameter}] than considering the bond structure factor $B(q)$. Within the Mott state,
instead, $B(\pi)$ is much reduced, still having a cusp. Let us now turn to density and spin correlations. For both insulating phases, $N(q)$ is smooth and shows a
quadratic behavior ($N(q) \propto q^2$) for small $q$, indicating that the charge excitations are gapped~\cite{capello2005}. Also the spin structure factor $S(q)$
in the Peierls phase is smooth, signaling that also spin excitations are gapped. By contrast, in the Mott phase $S(q)$ is linear for small $q$ and has a peak at
$q=\pi$, which is expected to diverge logarithmically with the system size and signals the presence of gapless spin excitations.

\begin{figure*}
\includegraphics[width=0.9\textwidth]{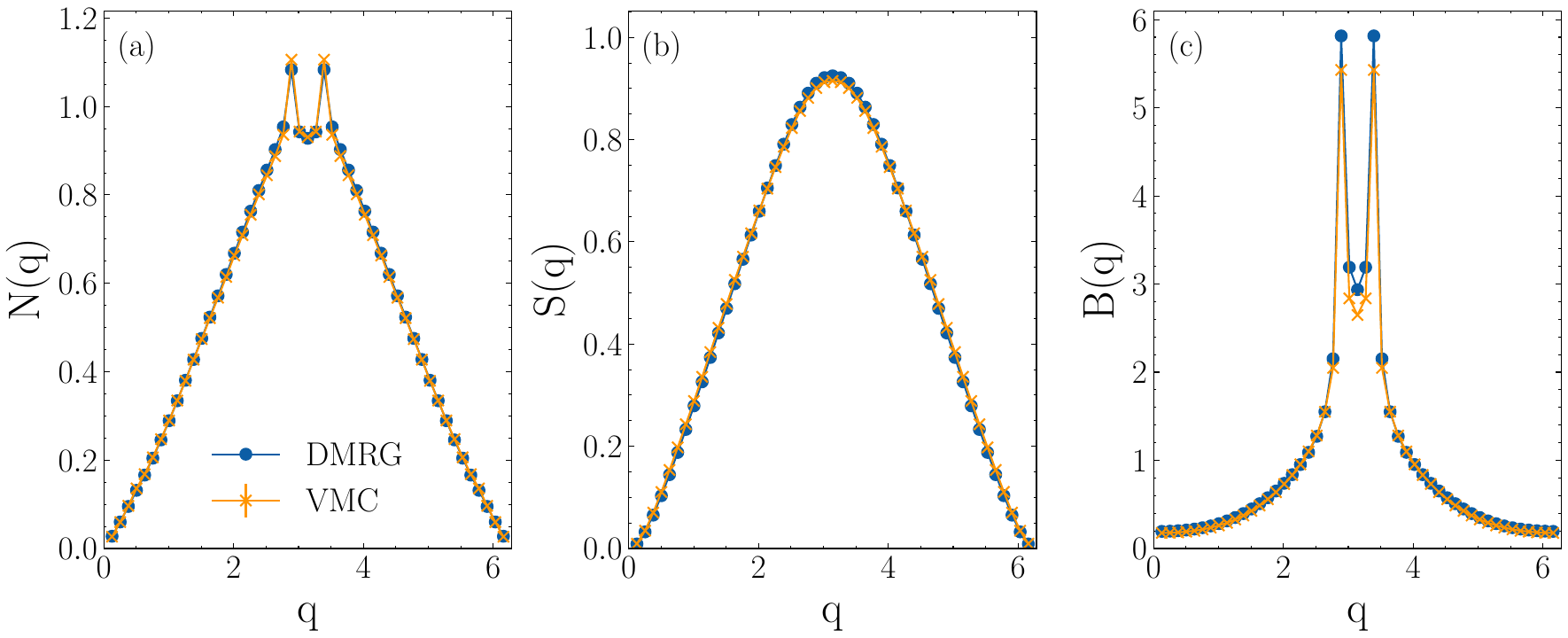}
\caption{\label{fig:doped_U0}
Density, spin, and bond structure factors for a chain with $L=50$ sites at $\delta=0.08$, for $U=0$ and $\lambda=0.16$ (with $\hbar \omega/t=1$). The results obtained
by using DMRG (optimized for the Hamiltonian with periodic-boundary conditions) are also shown for comparison.}
\end{figure*}

\subsection{Results away from half-filling}

We now move to the most important part of this work and examine the behavior of the ground state in the doped regime, with a number of electrons $N_{\rm e}<L$, and
a doping defined by $\delta=1-N_{\rm e}/L$. We restrict our attention to the subspace where the total spin projection $\sum_i \hat{S}^z_{i}=0$. Let us start by
doping the Mott insulator. In the Hubbard model, with only nearest-neighbor hopping $t$ and without phonons, a Luttinger liquid is obtained, with gapless modes in
both charge and spin sectors~\cite{giamarchi}. In turn, a power-law decay (in real space) and cusps at $q=2k_F$ (in momentum space), where $k_F=\pi(1-\delta)/2$,
are present in the ground-state correlations. These features are also observed in the presence of phonons, as long as the Mott insulator is stabilized at half-filling,
see Fig.~\ref{fig:doped_Mott}. Here, the cusps at $q=2k_F$ are clearly visible in the variational calculations of $N(q)$ and $S(q)$. We notice that cusps are less
evident in the density correlations within DMRG, especially when approaching half-filling. Although these singularities are less pronounced than in the backflow wave
function, the linear behavior of $N(q)$ for small values of $q$ (indicating gapless charge modes for $q \to 0$) leaves no doubt on the metallic nature of the system.
We notice that, in the doped case, DMRG calculations are much more demanding than in the half-filled regime; nevertheless, the variance of the total energy is
always lower than $0.09t^2$.

The most interesting outcome appears when doping the Peierls insulator. In this case, a Luther-Emery liquid emerges at small values of the doping. Indeed, for
$\delta>0$ the system immediately turns into a metal, but the spin gap remains finite close to half-filling. By further increasing $\delta$, the spin gap gradually
decreases, eventually leading to a Luttinger liquid for sufficiently large values of $\delta$. This fact is particularly interesting since the Luther-Emery state has
been found in repulsive models with no phonons, emerging from a spin-gapped and dimerized insulator~\cite{fabrizio1996,daul1998,nishimoto2008,balents1996,shen2023},
however, its stabilization has been argued to heavily rely on the existence of multiple Fermi points in the non-interacting band structure. By contrast, in the present
case the band structure is trivial, with only two Fermi points, and the spin gap is opened by the e-ph coupling. The results of the density and spin structure factors
are reported in Fig.~\ref{fig:doped_Peierls}, where a smooth $S(q)$ persists in the doped system, before the insurgence of cusps at $q=2k_F$. The existence of a
spin-gapped metal is confirmed by DMRG. Curiously, the Luther-Emery in the Hubbard-SSH model has not been detected before, even though its stability extends down to
the $U=0$ limit, where a finite e-ph interaction drives into a Peierls state at half-filling, see Fig.~\ref{fig:phasdiag}. The exemplification of this fact is shown
for $\lambda=0.16$ and $\delta=0.08$ in Fig.~\ref{fig:doped_U0}. Even though a precise determination of the power-law decay of correlation functions is difficult
(given their oscillatory behavior), we have evidence that the bond-bond correlations are the dominant ones in the Luther-Emery phase (not shown). 

\begin{figure}
\includegraphics[width=\columnwidth]{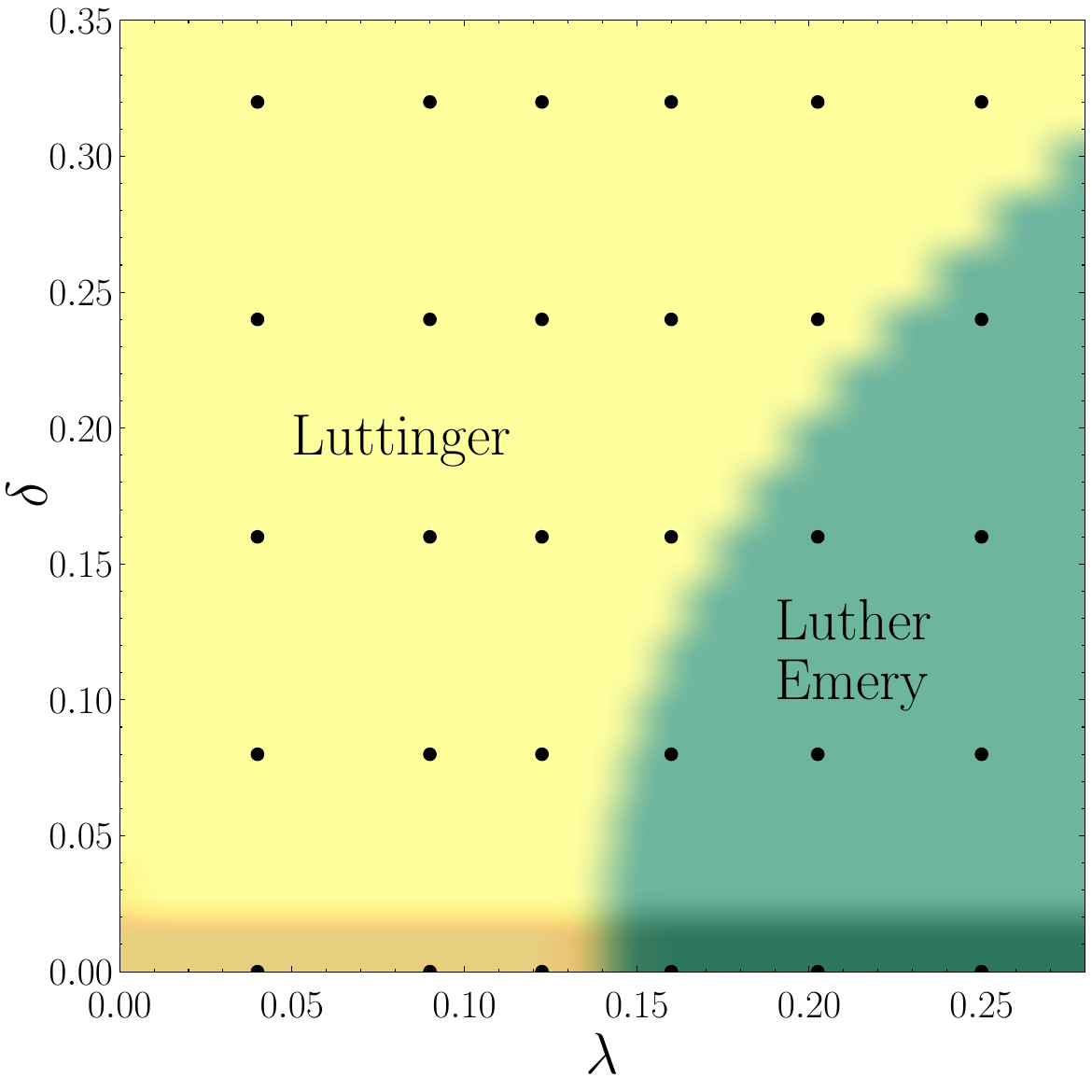}
\caption{\label{fig:phasediag_doped_U4}
Phase diagram at $U/t=4$ as a function of $\lambda$ and $\delta$ with $\hbar \omega/t=1$. Calculations are performed on a chain with $L=50$ sites, with data points represented by filled circles. The system is insulating at $\delta=0$, indicated by the shaded region at the bottom of the figure. For finite values of $\delta$, a transition is observed from a Luttinger liquid to a Luther-Emery metallic phase.}
\end{figure}

\begin{figure*}
\includegraphics[width=0.9\textwidth]{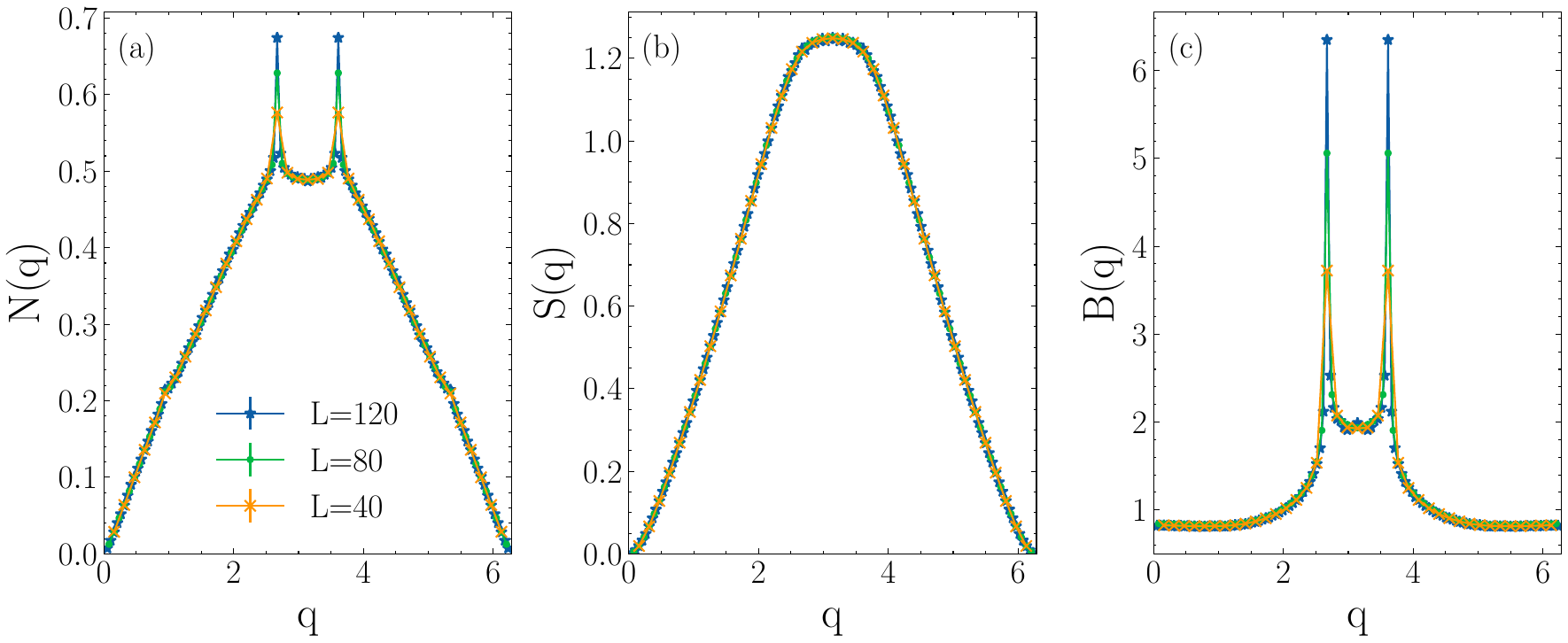}
\caption{\label{fig:size_scaling_LE}
Density, spin, and bond structure factors for a Luther-Emery metal at $\lambda=0.25$ and $\delta=0.15$; the quantities are shown for chains of lengths $L=40$, $80$,
and $120$. No significant changes are observed in the spin structure factors, confirming that the Luther-Emery liquid survives in the large-chain limit; the peaks
of the charge and bond structure factors at $2k_F$ diverge sub-linearly with the system size.}
\end{figure*}

By analyzing the presence of singularities in the spin structure factor, we are able to construct the phase diagram where the electron doping $\delta$ is varied.
In particular, we report the case in which $\lambda$ is varied and $U/t=4$ is kept fixed in Fig.~\ref{fig:phasediag_doped_U4}. Here, a Luther-Emery metal exists
when (slightly) doping the Peierls insulator at $\delta=0$ and sufficiently large values of $\lambda$. By further increasing the doping concentration, the ground
state becomes a  Luttinger liquid, with gapless spin excitations. To confirm the stability of the Luther-Emery phase and ensure that the spin gap persists as the
system size increases, we also perform a size-scaling analysis, focusing at $\lambda=0.25$ and $\delta=0.15$ (which falls within the Luther-Emery phase phase) and
display the structure factors for three different sizes (i.e., $L=40$, $80$, and $120$), see Fig.~\ref{fig:size_scaling_LE}. Remarkably, the spin structure factor
does not show substantial modifications, keeping a completely smooth behavior, which implies the stability of the Luther-Emery phase also on large systems (and
eventually in the thermodynamic limit). For $U/t=4$, we rule out the presence of phase separation in both the Luttinger and Luther-Emery phases within the range
of $\lambda$ and $\delta$ shown in Fig.~\ref{fig:phasediag_doped_U4}. This conclusion is clearly supported by DMRG simulations, which gives the correct (convex)
behavior of the energy as a function of the electron doping; VMC results show a tiny region of phase separation close to half filling in the Luther-Emery phase,
but this fact is primarily due to the different accuracy of VMC energies, which is very high in the Peierls insulator and slightly worse for small $\delta$ (still grasping the
correct physical behavior). 

\section{Conclusions}\label{sec:concl}

In conclusion, we showed that a variational {\it Ansatz}, which goes beyond the simple Jastrow-Slater approximation, can characterize the various phases of the
one-dimensional Hubbard-Su-Schrieffer-Heeger model both at half-filling and in the hole-doped regime. Our results are in good agreement with DMRG calculations,
confirming that the physical description of the system we are providing is accurate. In addition, we are able to convincingly show the presence of a spin-gapped
metallic phase upon doping the Peierls insulator. Since the variational wave function we propose can be readily generalized in two dimensions and variational Monte
Carlo methods do not suffer limitations from considering higher-dimensional models, our next step will be to study a two-dimensional Hubbard-Su-Schrieffer-Heeger
model within the same framework. We are particularly interested in the doped regime, to investigate how the presence of phonons can favor various types of
instabilities, such as stripe order or $d$-wave superconductivity\cite{wang2022}. We remark that a one-dimensional band structure possesses nesting at $2k_F$,
which makes the superconducting fluctuations compete against the bond-order ones, with the latter dominant in the Hubbard-Su-Schrieffer-Heeger model. In higher
dimensions, nesting is a rather exceptional circumstance, especially away from commensurate densities, suggesting that, should our findings extend beyond one
dimension, superconductivity would be the dominant instability channel.

\section*{Acknowledgements}
We thank L.L. Viteritti for his precious help regarding the JAX implementation of the variational Monte Carlo code. F.F. acknowledges support by the Deutsche Forschungsgemeinschaft (DFG, German Research Foundation) 
for funding through TRR 288 -- 422213477. We acknowledge the CINECA award under the ISCRA initiative, for the availability of high performance computing resources and support.

\pagebreak

\bibliography{references}
\end{document}